\documentclass[aps,prd,preprint,groupedaddress]{revtex4}
\usepackage{amssymb}
\usepackage{graphicx}
\usepackage{bm}
\usepackage{amsmath}

\begin{document}


\title{Approximate stress-energy tensor of the massless
spin-$1/2$ field in Schwarzschild spacetime}
\author{Jerzy Matyjasek}
\email{matyjase@tytan.umcs.lublin.pl, 
jurek@kft.umcs.lublin.pl}
\affiliation{Institute of Physics, 
Maria Curie-Sk\l odowska University\\
pl. Marii Curie-Sk\l odowskiej 1, 
20-031 Lublin, Poland}

\date{\today}

\begin{abstract}
The approximate stress-energy tensor of the conformally invariant
massless spin-$1/2$ field in the Hartle-Hawking state in the
Schwarzschild spacetime is constructed. 
It is shown that by solving the conservation equation in conformal
space and utilizing the regularity conditions in a physical metric
one obtains the stress-energy tensor that is in a good agreement
with the numerical calculations.
The back reaction of the
quantized field upon the spacetime metric is briefly discussed.
\end{abstract}


\maketitle


Recently, the renormalized stress-energy tensor of the quantized
conformally invariant massless spin-1/2 field in the Schwarzschild and
Reissner-Nordstr\"om spacetimes has been evaluated
numerically~\cite{Paul1+,Paul2+}. This important and long awaited
result completes our knowledge of the behaviour of three basic quantum
test fields in the Hartle-Hawking state. However, obtained as a by-
product of the calculations its analytical approximation is, contrary
to the spin 0 and spin 1 cases, very poor near the event horizon
(although it is accurate at large distances).

It has been shown~\cite{Howard,Bruce+} that the stress-energy tensor
of the massless fields in the Schwarzschild geometry decomposes naturally as
\begin{equation}
   \langle T_{\nu }^{\mu }\rangle_{ren}\,=\,\langle T_{\nu }^{\mu }
   \rangle^{{\rm analytic}}\,
   +\,\Delta_{\mu}^{\nu},
                                 \label{dcmp}
\end{equation}
where $\langle T_{\nu }^{\mu }\rangle^{{\rm analytic}}$ is the
analytical approximation of the stress-energy tensor and 
$\Delta_{\mu}^{\nu}$ 
is a traceless and covariantly conserved tensor that must be
calculated numerically. For the conformally invariant massless scalar
field the analytical part coincides with the Page
approximation~\cite{Page} (and the approximations constructed
in~\cite{BOP,FZ,jurek2}), whereas for the vector fields such
approximation has been constructed in Refs.~\cite{Bruce+,jurek2}. It
has been shown that for both scalar and vector fields the
approximation is reasonable. (Analytical approximation of tensor
$\Delta_{\mu}^{\nu}$ of the scalar field has been constructed 
in~\cite{jurek3}).

Unfortunately, as has been pointed out in Ref.~\cite{Paul2+},
$\langle T_{\nu }^{\mu}\rangle^{{\rm analytic}}$ for
the spin-1/2 field gives a wrong sign of the energy density at the
event horizon, invalidating thus any prospect applications.
On the other hand, however, the expectation value of the stress-energy
tensor of the conformally invariant massless  fields in the Hartle-
Hawking state in the Schwarzschild geometry is known to possess some
general features~\cite{SteveSq}. The asymptotic behavior of tangential
and radial components of $\langle T_{\nu }^{\mu }\rangle_{ren},$ the
regularity conditions on the event horizon and the trace anomaly are
quite restrictive, and allow construction of a class of approximate
tensors. Further, a piece of the numerical data, such as the exact 
value of one of the components of the stress-energy tensor, 
say $\langle T_{\theta}^{\theta }\rangle_{ren},$ on the
event horizon may be used in the final determination of the model. It
should be noted that calculations of the horizon value of the stress-
energy tensor could be regarded as a relatively simple task since it
is sufficient to retain only the two smallest frequencies in the mode
sums~\cite{Candelas,Paul2+}. In this sense this information could be
regarded as independent of the numerical calculations of the stress-
energy tensor for $r > 2M.$

The idea of reconstructing $\langle T_{\nu }^{\mu }\rangle_{ren}$ from
the knowledge of its asymptotic behaviour is not new and belongs to
Christensen and Fulling~\cite{SteveSq}. It has been subsequently
elaborated in Refs.~\cite{Tadaki,Vaz1,Vaz2,jurek1}. In this note we
shall show that, contrary to the widespread opinion, it is possible to
construct the stress-energy tensor of massless spin 1/2 field in the
Schwarzschild spacetime, which satisfactorily approximates the `exact'
$\langle T_{\theta}^{\theta }\rangle_{ren}.$ The method is similar to
that of Ref~\cite{jurek4} with the one reservation: here we consider
the four component spinors rather than the two component ones.
Moreover, we shall explicitly demonstrate that the Frolov-Zel'nikov
method~\cite{FZ} also yields reasonable results and subsequently
compare the stress-energy tensors constructed within the frameworks of
both methods.

Let us start by counting  available information. First, it is known
that the stress-energy tensor is covariantly conserved and its trace
is given by a general formula
\begin{equation}
\langle T^{\mu}_{\mu}\rangle_{ren}\,=\,\alpha \left( {\cal H} + \Box R \right)+ 
\beta {\cal G} + \gamma \Box R,
\end{equation}
where 
\begin{equation}
{\cal H} = R_{\mu \nu \rho \tau} R^{\mu \nu \rho \tau} 
- 2 R_{\mu \nu} R^{\mu \nu} +
\frac{1}{3}R^{2},
\end{equation}
\begin{equation}
{\cal G} = R_{\mu \nu \rho \tau} R^{\mu \nu \rho \tau} 
- 4 R_{\mu \nu} R^{\mu \nu} + R^{2}
\end{equation}
and the numerical coefficients for the spin-1/2 field are 
$\alpha\,=\,18\,\lambda,$ $\beta\,=\,-11\,\lambda,$  $\gamma\,=\,0,$ 
and $\lambda\,=\,2^{7}45\pi^{2}.$
Further, we observe  $\langle T^{\mu}_{\nu}\rangle_{ren}$ is regular
on the past and future event horizon and approaches at large distances
the flat spacetime radiation stress-energy tensor. Finally, we expect
that the curvature effects enter the stress-energy tensor as
$\sim\,x^{3}$ ($x = 2M/r$). This requirement is usually motivated by
the observation that far from the event horizon the stress-energy
tensor should consist of the red- shifted thermal bath part
supplemented by the quantum corrections. This assumption is sometimes
referred to as a weak thermal bath hypothesis~\cite{Visser}.

The idea is to construct $\tilde{T}^{\mu}_{\nu}$ in the
optical metric and subsequently to transform it back to the physical
spacetime with the aid of a transformation that relates the stress-
energy tensor in conformally related geometries~\cite{BO,BOP,Page}.
Now, the independent informations listed above suggest that the
tangential component of the stress-energy tensor could be approximated
as
\begin{equation}
{\tilde T}_{\theta}^{\theta}\,=\,\frac{7}{2}T\left(1\,+\,
\sum_{i}^{N}a_{i}x^{i}\right),
\end{equation}
with $N=6$ and $T^{-1}\,=\,90\pi^{2}(8M)^{4}.$

Returning to the Schwarzschild geometry and making use of the
regularity conditions in the physical metric, after some algebra, 
one has
\begin{equation}
 T_{\theta }^{\theta } \,=\,{\frac{7}{2}}T\left[
1\,+\,2x\,+\,3x^{2}\,+{\frac{68}{7}}x^{3}\,+\,{\frac{1}{14}}%
(230\,+\,14a_{4})x^{4}\,+\,{\frac{102}{7}}x^{5}\,+\,{\frac{87}{7}}%
x^{6}\right] .
                                       \label{1_parametr_c}
\end{equation}
 Remaining components of the approximate stress-energy tensor can be
easily obtained from (\ref{1_parametr_c}) and will not be displayed
here. Similar results for the two component spinors have been constructed
in Ref.~\cite{jurek4} to which the reader is referred for the
technical details. It should be noted that the logarithmic term
$
x^{2}\left(a_{1}\,+\,2 a_{2}\right) \ln x,
$
which appears as the result of integration of the conservation
equation survives even if the regularity conditions are satisfied. 
Only after accepting the weak thermal bath hypothesis the coefficients
$a_{1}$ and $a_{2}$  could be equated to zero~\cite{jurek2}.
The parameter $a_{4}$ can be determined form the equation:
\begin{equation}
T_{\mu}^{\nu}(2M)\,=\,\langle T_{\mu}^{\nu}(2M)\rangle ,
                                   \label{EH}
\end{equation}
which, by the spherical symmetry, equality of $T_{t}^{t}$ and $T_{r}
^{r}$ at the event horizon and the trace anomaly yields in fact only
one condition. 

Although the horizon values of the components of the
stress-energy tensor in the Hartle-Hawking state have never been cited
explicitly in literature, and Carlson et al.~\cite{Paul2+} present
their results only graphically, we have sufficient informations to
construct the model. The horizon value of the trace anomaly in the
case at hand is $7/7680\pi^{2}M^{4},$
whereas the approximate values of  $T_{\theta}^{\theta}$ is~\cite{Paul2+}
\begin{equation}
T_{\theta}^{\theta}(2M)\,=\,T_{\phi}^{\phi}(2M)\,\approx\,
\frac{105}{90\pi^{2}(8M)^{4}}.
                                                  \label{ang}
\end{equation}
Making use of Eqs.~(\ref{1_parametr_c}) and Eq.~(\ref{EH}) 
with the right hand side
given by (\ref{ang}) one finally obtains 
\begin{equation}
T_{t}^{t}\,=\,-\frac{7}{2}T\left( 3 + 6\,x + 9\,x^{2} + 12\,x^{3}
- \frac{39\,x^{4}}{7} + \frac{186\,x^{5}}{7} - 
  69\,x^{6}\right),
                                                  \label{appr_tt}
\end{equation}
\begin{equation}
T_{r}^{r}\,=\,\frac{7}{2}T\left(  1 + 2\,x + 3\,x^2 - 
\frac{52\,x^3}{7} + \frac{139\,x^4}{7} -
  \frac{18\,x^5}{7} + \frac{15\,x^6}{7}\right)
                                                 \label{appr_rr}
\end{equation}
and
\begin{equation}
T_{\theta}^{\theta}\,=\,\frac{7}{2}T\left(  1 + 2\,x + 3\,x^2 + 
\frac{68\,x^3}{7} - \frac{89\,x^4}{7} + 
  \frac{102\,x^5}{7} + \frac{87\,x^6}{7}\right).  
                                                \label{appr_ang}
\end{equation}
The run of the components of  $T_{\mu}^{\nu}$ are displayed in Figs. 1-3.
The new `radial' coordinate $\xi$ is defined as $\xi=(r-r_{+})/M,$
where $r_{+}$ denotes location of the event horizon.

It should be emphasized that the case of the spin-$1/2$ fields is
special. Indeed, for the electromagnetic field the $T_{\mu}^{\nu}$
coincides with the analytic part of 
$\langle T_{\mu}^{\nu}\rangle$~\cite{jurek2},
whereas the analogous result constructed for the conformally invariant
scalar fields substantially improves the Page approximation.

A different method of calculating the stress-energy tensor in static
spacetimes has been proposed by Frolov and Zel'nikov in
Ref.~\cite{FZ}. It has been shown that it is possible to construct a
family of expressions describing  the approximate stress-energy tensor
$T_{\mu}^{(FZ)\nu}$ solely form the curvature, the Killing vector and
their covariant derivatives up to some given order. By construction
the Frolov-Zel'nikov tensors have a correct trace and a proper
behaviour under the scale transformations. Upon imposing appropriate
regularity conditions at the event horizon of the Schwarzschild black
hole one obtains a one parameter family of the approximate stress-
energy tensor. As expected, the resulting expressions are simple
polynomials in $x.$ If for the conformally invariant scalar fields the
free parameter is set to zero the result  coincides with the Page
approximation. It should be noted, however, that there are no {\it
a~priori} reasons to accept such a choice. Indeed, for a vector field
such a choice leads to evidently wrong results. On the other hand the
free parameter may be easily adjusted employing Eq.~(\ref{EH}).

It could be easily shown that
\begin{equation}
T^{(FZ)\nu}_{\mu}\,=\,T_{\mu}^{\nu}\,+\,D_{\mu}^{\nu},
                                                 \label{FZ}
\end{equation}
where $T_{\mu}^{\nu}$ is the approximate stress-energy tensor 
(\ref{appr_tt}-\ref{appr_ang}) and the conserved and
traceless tensor $D_{\mu}^{\nu}$ is given by
\begin{equation}
D_{t}^{t}\,=\,-\frac{7}{2}T\left( \frac{783\,x^4}{35} - 
\frac{174\,x^5}{35} - \frac{87\,x^6}{5}\right),
                                               \label{dFZtt}
\end{equation}
\begin{equation}
D_{r}^{r}\,=\,\frac{7}{2}T\left(\frac{232\,x^3}{35} - 
\frac{667\,x^4}{35} + \frac{174\,x^5}{35} + \frac{261\,x^6}{35}\right)
                                              \label{dFZrr}
\end{equation}
and
\begin{equation}
D_{\theta}^{\theta}\,=\,-\frac{7}{2}T\left(\frac{116\,x^3}{35} - 
\frac{145\,x^4}{7} + \frac{174\,x^5}{35} +   \frac{87\,x^6}{7}\right).
                                              \label{dFZang}
\end{equation}
The components of $D_{\mu}^{\nu}$ are displayed in Figs. 1-3 (right
panel). 

Now we can compare the approximate tensors
(\ref{appr_tt}-\ref{appr_ang}) to (\ref{FZ}-\ref{dFZang}) and to the
numerical results presented in Ref.~\cite{Paul2+}. By construction the
tensors  are exact at infinity and very close to the exact value at
the event horizon. Inspection of similar figures presented in
Ref.~\cite{Paul2+} shows that our approximation
(\ref{appr_tt}-\ref{appr_ang}) is in good agreement with the exact
numerical calculations. The Frolov-Zel'nikov approximation is slightly
worse but still reasonable. Unfortunately, we are unable to provide
detailed comparison between the exact numerical results and the
approximations as the former were presented only graphically. 

The run of $T_{\mu}^{\nu}$ and $T_{\mu}^{(FZ)\nu}$ constructed 
for the massless spinor field qualitatively resembles
behaviour of the approximate stress-energy tensor of the
electromagnetic field~\cite{Bruce+}. Indeed, for the electromagnetic
field the Frolov-Zel'nikov
approximation is also less accurate than the analytic part of $\langle
T_{\mu}^{\nu}\rangle .$

\begin{figure}[th]
\includegraphics[height=6cm]{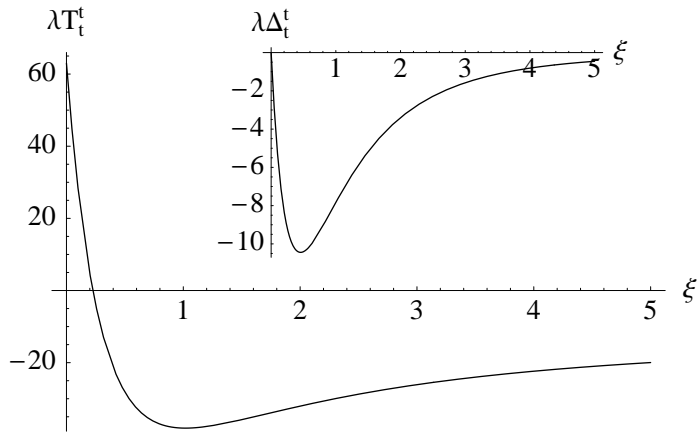}
\caption{This graph shows the radial dependence of the rescaled
component $T_{t}^{t}$ $[\lambda = 90 (8M)^4\pi^2 ]$ of the approximate stress-energy
tensor of the massless spin 1/2 field in Schwarzschild spactime and
the rescaled component $D_{t}^{t}$ (right panel).\label{figa}}
\end{figure}

\begin{figure}[th]
\includegraphics[height=6cm]{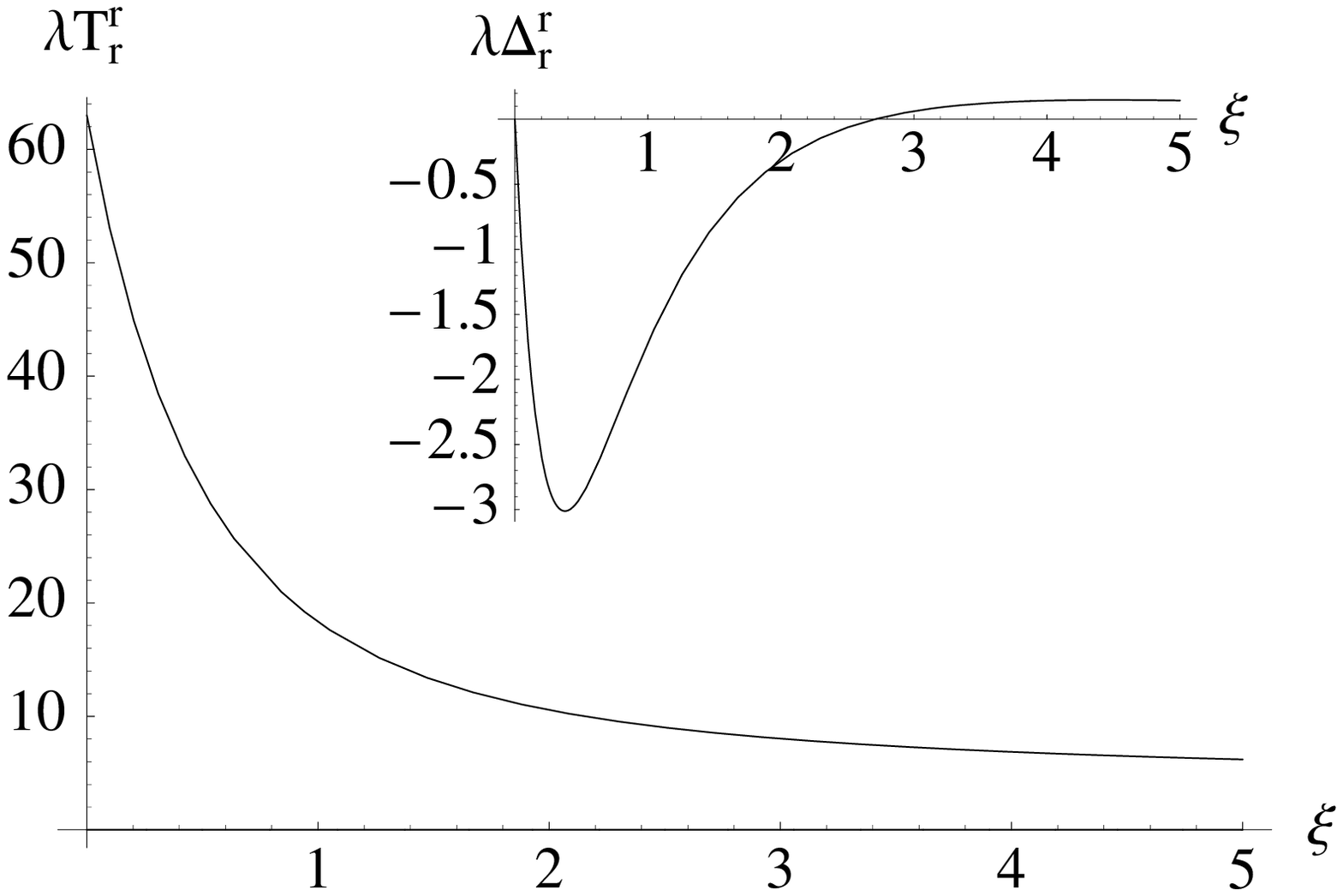}
\caption{This graph shows the radial dependence of the rescaled
component $T_{r}^{r}$ $[\lambda =  90 (8M)^4\pi^2  ]$ of the approximate stress-energy
tensor of the massless spin 1/2 field in Schwarzschild spactime and
the rescaled component $D_{r}^{r}$ (right panel).\label{figb}}
\end{figure}

\begin{figure}[th]
\includegraphics[height=6cm]{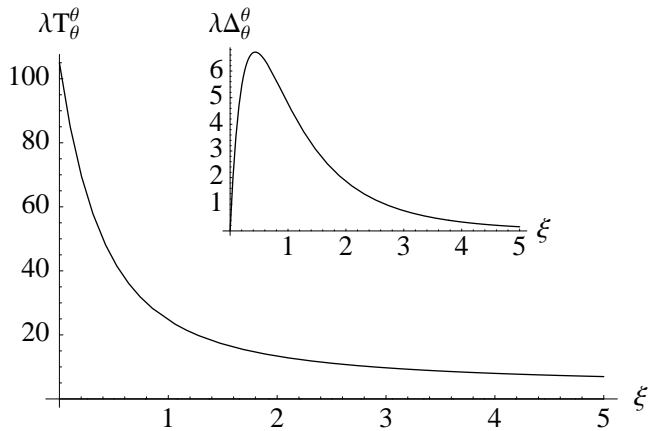}
\caption{This graph shows the radial dependence of the rescaled
component $T_{\theta}^{\theta}$ $[\lambda =  90 (8M)^4\pi^2 ]$ of the approximate stress-energy
tensor of the massless spin 1/2 field in Schwarzschild spactime and
the rescaled component $D_{\theta}^{\theta}$ (right panel).
\label{figc}}
\end{figure}

Now we shall employ the constructed approximation in the backreaction calculations.
It should be noted however, that in order to investigate 
the evolution of the system consisting of a
black hole and the quantized field it is necessary to have at one's
disposal the detailed knowledge of the functional dependence of the
renormalized stress-energy tensor on a wide class of metrics. So long
the effects of the quantum gravity could be safely ignored this is
achieved solving the semi-classical equations:
\begin{equation}
G_{\mu}^{\nu}[g]\,=\,8\pi \left(T_{\mu}^{(cl)\nu}[g] + 
\langle T_{\mu}^{\nu}[g]\rangle\right) .
                                                   \label{back}
\end{equation}
The first term in the right hand side of the above equation comes from
the classical source whereas the second one is due to the contribution of the
quantized fields.
Unfortunately, all our current understanding of the stress-energy
tensor of the quantized massless fields is limited to the static black
holes described by the electric charge and the mass.
Our results, however, may be of use in the
linearized back reaction calculations~\cite{York!,Lousto+}: the curved 
spacetime leads to the
stress-energy tensor approximated by (\ref{appr_tt}-\ref{appr_ang}), 
which, in turn, modifies the background geometry.

Since the stress-energy tensor is asymptotically constant it is
necessary to put the system in the cavity or spherical box of definite
radius, $r_{c}.$ It should be noted, that presence of the boundary
certainly modifies the stress-energy tensor. Therefore the radius $r_{c}$
should be chosen to guarantee applicability of the perturbative approach
on the one hand and to minimize the error caused by
ignoring expected boundary effects on the other.

The geometry of the quantum corrected spherically-symmetric static 
black hole is generally described by the
line element
\begin{equation}
ds^{2}\,=\,-e^{2\psi(r)}f(r) dt^{2}\,+\,
\frac{1}{f(r)}
dr^{2}\,+\,r^{2}d\Omega^{2},
                               \label{line_el}
\end{equation}
where $f(r) = 1- 2m(r)/r $ and $d\Omega^{2}$ is the metric on a unit 
sphere. We shall assume that the  functions $m(r)$  and  $\psi(r)$  
can be expanded as
\begin{equation}
m(r)\,=\,M\,+\,\varepsilon M_{1}(r)\,+\,{\cal O}(\varepsilon^{2})
                               \label{mr}
\end{equation}
and
\begin{equation}
\psi(r)\,=\,\varepsilon \psi_{1}(r)\,+\,{\cal O}(\varepsilon^{2}).
                              \label{psir}
\end{equation}
To keep control of the order of terms we have introduced the 
dimensionless parameter $\varepsilon,$ which will be set to 1  
at the final stage of calculations. Similarly, 
$\langle T_{\mu}^{\nu}[g]\rangle $ in the right hand side of~Eq.(\ref{back})
should be substituted by 
$\varepsilon \langle T_{\mu}^{\nu}[g]\rangle .$

The solutions of the linearized semi-classical Einstein field
equations with a source term given by the stress-energy tensor 
(\ref{appr_tt}-\ref{appr_ang}) reduce to
two elementary quadratures:
\begin{equation}
M_{1}(r)\,=\,-4\pi \int r^{2} T_{t}^{t} dr\,+\,C_{1}
                                       \label{M1}
\end{equation}
and
\begin{equation}
\psi_{1}(r)\,=\,4\pi \int \frac{r^{2}\left(T_{r}^{r} - 
T_{t}^{t}\right)}{r - 2M}\,+\,C_{2},
                                      \label{psi1}
\end{equation}
where the integration constants $C_{1}$ and $C_{2}$ are to be
determined form the boundary conditions. 
Observe that it is possible to determine the function $\psi_{1}(r)$
without prior knowledge of the horizon value of the stress-energy tensor.
It is simply because the difference $T_{r}^{r} - T_{t}^{t}$ 
does not depend on the coefficient
$a_{4}.$ 

Our preferred choice of the boundary condition for
Eq.~(\ref{M1})
is simply $M_{1}(r_{+})\,=\,0,$
which requires knowledge of the
exact location of the event horizon, $r_{+}.$
From (\ref{mr}) and (\ref{M1}) one has $r_{+} = 2 M,$ and hence $M$ is to be 
interpreted as the horizon defined mass.
For the function $\psi_{1}\left(  r\right)$ we shall adopt
the natural condition
$g_{tt}(r_{c}) g_{rr}(r_{c})\,=\,-1.$

Now the equations (\ref{M1}) and (\ref{psi1}) can be easily integrated to yield
\begin{eqnarray}
M_{1}\left(  r\right) &=&\frac{K}{M}\left(  \frac{7}{6\,x^{3}}+\frac
{7}{2\,x^{2}}+\frac{21}{2\,x}-33+\frac{13}{2}x-\frac{31}{2}x^{2}\right. 
\nonumber \\
&&\left. +\frac{161}
{6}x^{3}-14\,\ln x\right)
                             \label{M1sol}
\end{eqnarray}
and
\begin{eqnarray}
&&\psi_{1}\left(  r\right)  =\frac{K}{M^{2}}\left[  \frac{7}{6}\left(
\frac{1}{x^{2}}-\frac{1}{x_{c}^{2}}\right)  +7\left(  \frac{1}{x}-\frac
{1}{x_{c}}\right)-\frac{50}{3}\left(  x-x_{c}\right) \right. \nonumber \\
&&\left.  -\frac{25}{2}\left(
x^{2}-x_{c}^{2}\right)  -13\left(  \,x^{3}-x_{c}^{3}\right)  -14\,\ln\frac
{x}{x_{c}}\right],
                           \label{psi1sol}
\end{eqnarray}
where $K = 3840 \pi$ and $x_{c} = 2 M/r_{c}.$
It is believed that the line element (\ref{line_el}) with the metric potentials
given by Eqs.~(\ref{mr},\ref{psir}) with (\ref{M1sol}) and (\ref{psi1sol}) 
respectively better approximates the  physical
reality then the original (Schwarzschild) one.
Having the quantum corrected geometry of the black hole one may 
study its properties such as Hawking temperature, 
trace anomaly and its influence
on the motion of test particles. 
Since the calculations of these effects are elementary we shall not
dwell on them here.
We only remark that in view of the results of Ref.~\cite{Paul2+}
presented method is the only one, which is able to provide simple and
reasonable approximations to the exact stress-energy tensor. This
model may be thought of as a minimal one, i. e. requiring only one
numerical information: the horizon value of the stress-energy tensor,
which, as has been pointed out, is to certain extend independent of
the numerical calculations for $r > 2M.$ Of course, more complicated
models which approximate  the exact $\langle T_{\mu}^{\nu}\rangle $
better could be easily devised at the expense of the additional
numerical informations. 

Similar calculations with different
asymptotics may be used in construction of the stress- energy tensor of
the spin-$1/2$ field in the Unruh state. We expect that the
approximation will be reasonable and the results presented in
Refs.~\cite{jurek2,jurek5} suggest that it is a safe anticipation.


\end{document}